\documentclass{article}

\usepackage{amssymb}
\usepackage{amsmath}
\usepackage{multirow}
\usepackage{dsfont}
\usepackage{graphicx}
\usepackage{rotating}
\usepackage{sidecap}

\begin{document}

\title{The virus of my virus is my friend: ecological effects of virophage with alternative modes of coinfection}


\author{Bradford P. Taylor$^1$, Michael H. Cortez$^2$, Joshua S. Weitz$^{2,1,3}$\\
$^1$ School of Physics,
$^2$ School of Biology,\\
Georgia Institute of Technology, Atlanta, GA, USA\\
$^3$Corresponding author: jsweitz@gatech.edu}

\maketitle

\begin{abstract}
Virophages are viruses that rely on the replication machinery of other viruses to reproduce within eukaryotic hosts.  
Two different modes of coinfection have been posited based on experimental observation.  
In one mode, the virophage and virus enter the host independently.  
In the other mode, the virophage adheres to the virus so both virophage and virus enter the host together.  
Here we ask: what are the ecological effects of these different modes of coinfection?  
In particular, what ecological effects are common to both infection modes, and what are the differences particular to each mode?  We develop a pair of biophysically motivated ODE models of viral-host population dynamics, corresponding to dynamics arising from each mode of infection.  We find both modes of coinfection allow for the coexistence of the virophage, virus, and host either at a stable fixed point or through cyclical dynamics.  In both models, virophage tend to be the most abundant population and their presence always reduces the viral abundance and increases the host abundance.  
However, we do find qualitative differences between models.  
For example, via extensive sampling of biologically relevant parameter space, we only observe bistability when the virophage and virus enter the host together.  
We discuss how such differences may be leveraged to help identify modes of infection in natural environments from population level data.

\end{abstract}

\section{Introduction}

Virophages are recently discovered viruses of viruses \cite{lascola2008, fischer2011}.  To reproduce, a virophage must infect a eukaryotic host that is also infected by a larger virus \cite{lascola2008}.  These larger viruses, hereafter referred to as viruses, are classified as NucleoCytoplasmic Large DNA Viruses and have comparable physical sizes and genome lengths to small bacteria \cite{vanetten2010}.  These larger viruses require the host to reproduce; however, their relatively large genomes encode for their own transcriptional machinery and part of their translational machinery, allowing the virus to reproduce within the host in a cytoplasmic structure of viral origin termed the \textquotedblleft viral factory" \cite{mutsafi2010}.  When also present, the virophage is thought to utilize the transcriptional machinery of viral origin and reproduce within the viral factory.  The virophage genome is much smaller than the genome of the virus and does not encode for any of the constituent parts of the viral factory.  Hence, virophages reproduce obligately through coinfection.  The virophage serves a parasitic role to the virus as viral burst sizes are greatly reduced during coinfection \cite{lascola2008, desnues2010}.


Virophages are continually being discovered and appear to be widespread biological entities in clinical and environmental settings.  The first discovered virophage, termed Sputnik, was isolated from a virus, mamavirus, that was extracted from the water in cooling towers in Paris, France \cite{lascola2008}.  A later discovered strain of Sputnik, termed Sputnik2, is associated with mamavirus-like Lentillevirus and shares the host \textit{Acanthamoeba polyphaga}, which is a causative agent of the human eye disease keratitis \cite{cohen2011}.  More recently, a third strain of Sputnik was discovered along with evidence that all strains could associate with many more viral strains than previously thought \cite{gaia2013}.  A different but related virophage, termed Organic Lake Virophage (OLV), was discovered from environmental sequencing data obtained from a hyper-salinic Antarctic lake and is associated with an algal host that undergoes yearly bloom cycles \cite{yau2011}.  The first discovered marine virophage, Mavirus, is associated with the bacterivorous host, \textit{Cafeteria roenbergensis}, which is endemic among the global oceans \cite{fischer2011}.  These last two examples suggest that newly discovered virophages may have global implications on algal blooms and nutrient cycles \cite{ogataclaverie2008}.  In fact, a genomic study suggests that undiscovered virophages exist in dozens of more locations including at different depths in oceans and lakes across the globe and within humans and other animals \cite{zhou2013}.

Among the discovered virophages, two different primary means for coinfection seem plausible.  In one mode, which we call the independent entry mode, the virophage and virus independently enter the host cell.  In the other mode, which we term the paired entry mode, the virophage entangles with the virus and coinfection occurs when the composite enters the host.  The paired entry mode of coinfection is thought to be the utilized by Sputnik strains \cite{desnues2010}.  In addition, the paired-entry mode of coinfection, to our knowledge, has strong indirect support, e.g., images show virophage grouped around viruses during viral production suggesting an affinity \cite{desnues2010}.  Images of virophage and virus present in the same phagocytic vacuole after coinfection serve as further evidence \cite{desnues2010}.  Additionally, there is a hypothesized structural basis for virophage-virus entanglement.  The mamavirus, a virus associated with Sputnik strains, is coated with long, tendrils that likely function to induce phagocytosis.  Experimental tests suggest these fibers are coated with peptidoglycan.  This coating is hypothesized to promote viral mimicry of the bacterial prey of the host amoeba \cite{xiao2009}.  Mushroom-like fibers coat the capsid exterior of the virophage Sputnik \cite{sun2010}.  The function of these fibers are unknown, but it is hypothesized they interact with mamavirus fibers to promote associating into a composite \cite{desnues2010}.  In accordance with this hypothesis, Sputnik is unable to reproduce in mixed cultures with bald forms of mamavirus-like strains \cite{boyer2011}.

\begin{figure}[h!]
    \includegraphics[width=\textwidth]{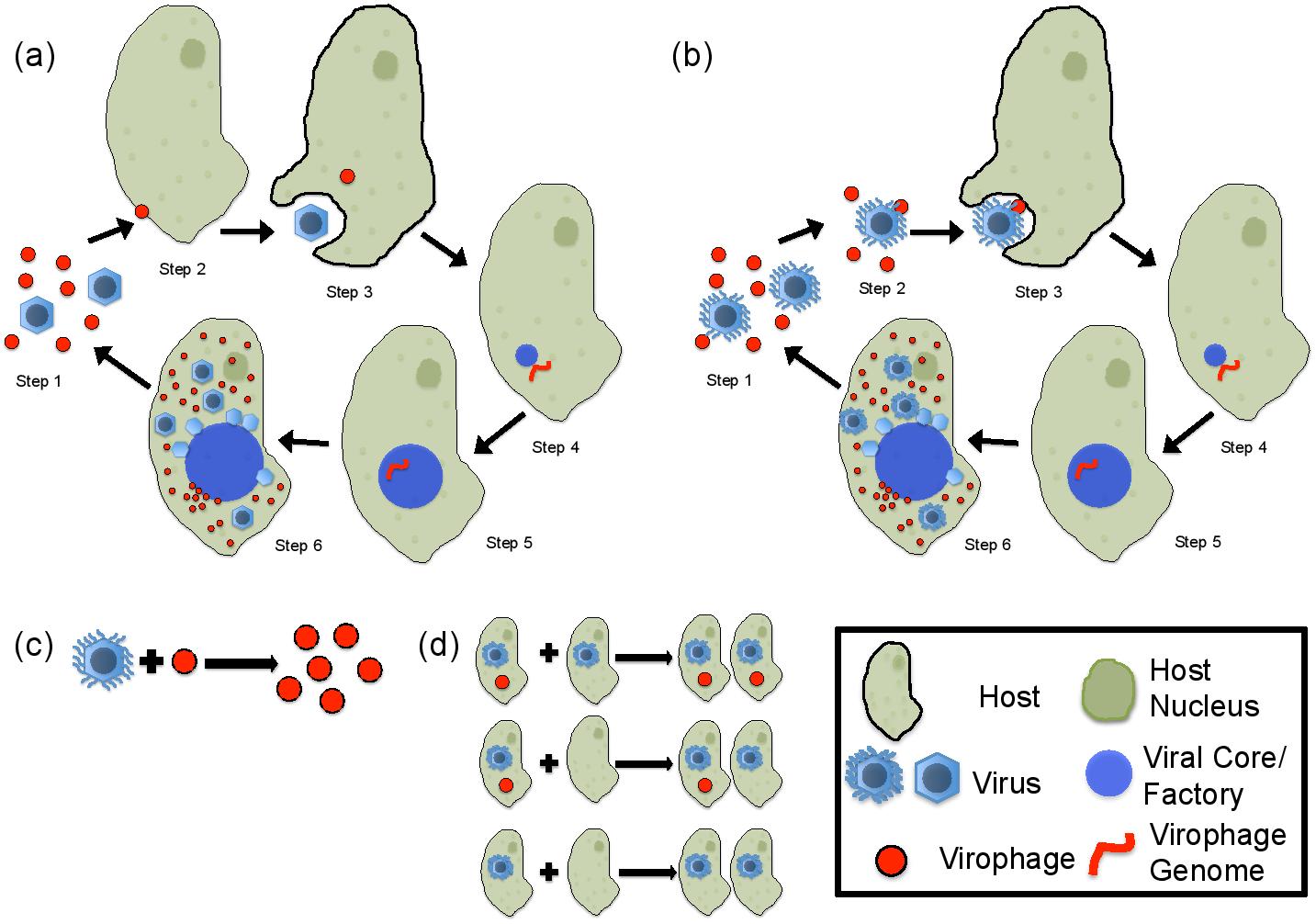}
      \caption{Stages during the virophage coinfection cycle in alternative mathematical models. (a) Independent entry mode, developed here: Step 1: free virus and virophage in the environment following host lysis.  Step 2: free virophage in the environment enters the host, note the host nucleus is shown as an internal large, darker green circle.  Step 3: free virus enters a host that previously engulfed a virophage.  Step 4: the viral particles lose their capsids. Step 5: the virophage genome enters the viral factory (large blue circle) which expands as viral particle genome replication occurs internally.  Step 6: fully formed viral particles bud from the viral factory and remain in the host cytoplasm until host lysis occurs.  (b) Paired entry mode, developed here, only steps 2 and 3 differ from the previous model.  Step 2: virophage attaches to virus to form a composite in the environment.  Step 3: the composite enters the host causing coinfection.  (c) Mechanism of virophage reproduction from a previous model: virophage reproduce via the infection and lysis of the virus in the absence of a host \cite{yau2011}. (d) Reproduction of virophage and virus in a direct contact model where free viral particles in the environment are not modeled \cite{wodarz2013}.\label{fig:flowchart}}
\end{figure}

The two modes for coinfection are pieces of a larger virophage infection process.  The entire coinfection processes for the independent entry and the paired entry modes are shown in Figures \ref{fig:flowchart}a and \ref{fig:flowchart}b, respectively.  The post-coinfection dynamics are considered equivalent between modes (steps 4-6).  Here, the viral core and virophage genome (of the virus and virophage, respectively), separate from their capsids (step 4).  Hereafter, we refer to viruses and virophage collectively as viral particles.  Note that this step has been experimentally observed for the virus but not for the virophage \cite{mutsafi2010}.  The viral factory originates from the viral core, which contains the viral genome.  The virophage genome enters the viral factory and the viral factory grows in size as replication of viral particle genetic material initiates inside (step 5) \cite{desnues2012b, mutsafi2010}.  Capsids form on the exterior of the viral factory and fully formed viral particles remain in the host cytoplasm until host lysis occurs.  Lysis typically occurs at about $16$ hours post-infection for Sputnik \cite{desnues2010}.

In this paper, we use theoretical models to explore how the ecological dynamics of the host, virus, and virophage populations depend on the biophysical mechanism of coinfection.  Our models correspond to the independent entry and paired entry mechanisms above.  A particularly important mechanistic difference between our models and two other models of virophage dynamics in the literature, is that we explicitly model the population dynamics of the viral particles in the environment.  
One model treated the virophage as a predator of the virus and modeled virophage growth as host independent (illustrated in Figure \ref{fig:flowchart}c) \cite{yau2011}.  However, virophages require both host and virus for reproduction.  
The other model borrows from epidemiological theory by modeling the spread of viruses and virophage through direct-contact between hosts, i.e., it does not model free virus or virophage in the environment (illustrated in Figure \ref{fig:flowchart}d) \cite{wodarz2013}.  
We note that infection dynamics from models of direct and indirect disease transmission can coincide when viral dynamics (e.g., degradation) in the environment are very fast \cite{cortez2013}; however, we are unaware of experimental evidence to suggest this is the case.

In the rest of the paper, we first present our mathematical models for each mode of coinfection.  Next, we demonstrate that stable and cyclical coexistence occurs between the virus, virophage, and host in each mode.  In both models, we find that virophage coexistence results in a reduction of viral abundance and an increase in host abundance.  We then derive an effective theory of host-viral interactions that accounts for this virophage-mediated shift in population levels.  Finally, we identify differences in coexistence between two modes that may be leveraged in future efforts to identify the infection mode from population level data.

\section{Methods}
\subsection{General modeling framework}
For both modes of coinfection, we explicitly model the density of viruses and virophage in the environment (step 1 in Figures \ref{fig:flowchart}a,b).  The units for these densities are ml$^{-1}$.  Intracellular dynamics (lysis) are assumed to be instantaneous once infection occurs.  This is akin to step 3 pointing directly to step 1.  We assume a well-mixed system and assume the rates of contact follow mass action kinetics.  The host and viral dynamics are modeled from an adapted Lotka-Volterra framework.  Further assumptions specific to each model are discussed below.  Note that consideration of delays between infection and lysis have been previously analyzed in the viral modeling literature and would be an important area of future exploration in this context \cite{beretta2001, levin1977}.

\subsubsection{IEM:  independent entry model of virophage and virus}


In the independent entry model (IEM), the virus and virophage independently enter the host.  Coinfection occurs when a virus enters a host in which the virophage previously entered.  We model the dynamics between the host ($H$), virus ($V$), virophage ($P$), and the host with an internal form of the virophage ($H_p$), hereafter referred to as an infected host, see Figure \ref{fig:flowchart}a, steps 1-3.  The full model is:
\begin{eqnarray}\label{eqn:IEM}\begin{split}
\frac{dH}{dt} = &\overbrace{H\left[b-d\left(1+\frac{H+H_p}{K}\right)\right] + (1-\rho) b H_p}^{\text{host growth}}-\overbrace{(\phi_p P + \phi_v V)H,}^{\text{infection (virophage and virus)}}& \\
\frac{dH_p}{dt} = &\overbrace{H_p\left[\rho b - d\left(1+\frac{H+H_p}{K}\right)\right]}^{\text{infected host growth}} +\overbrace{\phi_p PH -\phi_v V H_p,}^{\text{infection (virophage and virus)}}&\\
\frac{dV}{dt} = &\overbrace{(\beta_v H +\beta_{vp} H_p) \phi_v V}^{\text{virus production (lysis)}} -\overbrace{m_v V,}^{\text{virus decay}}&\\
\frac{dP}{dt} = &\overbrace{\beta_p \phi_v V H_p}^{\text{virophage production (lysis)}}-\overbrace{\phi_p PH}^{\text{virophage infection}} -\overbrace{m_p P,}^{\text{virophage decay}}&
\end{split}\end{eqnarray}
where $b$ and $d$ are the density-independent host birth and death rates, respectively, and $\rho$ is the fraction of infected host offspring that remain infected after reproduction. $K$ is the host density at which the total death rate is twice that of the intrinsic death rate, $\phi_v$ is the absorption rate between virus and host, and $\phi_p$ is the host absorption rate of virophage.  The virus and virophage decay with rates $m_v$ and $m_p$, respectively.  The burst size of the virus is $\beta_{vp}$ during virophage coinfection and $\beta_{v}$ otherwise, while $\beta_p$ is the burst size for virophage.  We assume virophage and virus burst sizes during coinfection are linearly dependent on $\beta_v$ such that $\beta_p = \rho_p \beta_v$ and $\beta_{vp} = \rho_{vp} \beta_v$.  We assume virophage can not enter a host with a virophage already present.  We assume the virophage does not decay within the host.  We also assume the host pays no cost and gains no direct benefit while carrying the virophage.

\subsubsection{PEM:  paired entry model of virophage and virus}
In the paired entry model (PEM), coinfection occurs when a virophage attaches to a virus in the environment and the virophage-virus composite later enters the host.  We model the population dynamics of the host ($H$), virus ($V$), virophage ($P$), and the virophage-virus composite ($V_p$), as shown in steps 1-3 of figure \ref{fig:flowchart}b. The full model is:

\begin{eqnarray}\label{eqn:PEM}\begin{split}
\frac{dH}{dt} = &\overbrace{H\left[b-d\left(1+\frac{H}{K}\right)\right] }^{\text{host growth}}-\overbrace{\phi_v(V+V_p)H,}^{\text{infection (virus and composite)}}&\\
\frac{dV}{dt} = &\overbrace{(\beta_v V +\beta_{vp} V_p) \phi_v H}^{\text{virus production (lysis)}} -\overbrace{\phi_{vp} VP}^{\text{virophage adhesion}} +\overbrace{m_p V_p}^{\text{virophage decay}} -\overbrace{m_v V,}^{\text{virus decay}}&\\
\frac{dV_p}{dt} = &\overbrace{\phi_{vp} V P}^{\text{composite formation}} +\overbrace{\beta_i \phi_v V_p H}^{\text{composite burst (lysis)}}-\overbrace{(m_p+m_v) V_p,}^{\text{decay (virus and virophage)}}  \\
\frac{dP}{dt} = &\overbrace{\beta_p \phi_v V_p H}^{\text{virophage production (lysis)}} -\overbrace{\phi_{vp} VP}^{\text{composite formation}}+\overbrace{m_v V_p}^{\text{virus decay}} -\overbrace{m_p P,}^{\text{virophage decay}}&
\end{split}\end{eqnarray}
where $\phi_{vp}$ is the rate of entanglement between virus and virophage and $\beta_i$ is the virophage-virus composite burst size.  We assume the composite burst size is linearly dependent on $\beta_v$ such that $\beta_i = \rho_i \beta_v$.  The rest of the parameters  have the same meaning as in the IEM.  We assume only one virophage can entangle with a virus and, once entangled, either the virus or virophage can independently decay leaving the virophage or virus free, respectively.  We assume that some viruses and virophage emerge as composites after lysis.  We do not include the incorporation of the virophage (provirophage) into the viral DNA as observed between one strain of the virophage Sputnik-2 and one strain of the virus Lentillevirus \cite{desnues2012}.

\subsection{Biophysical parameters}

We obtained reference values for the model parameters either from the literature, from derivations based on first principles, or through personal communication (with Matthias Fischer).  We used the Mavirus virophage system as a reference for our IEM parameters.  Mavirus has been observed independently entering the host; however, the exact mechanism is not well understood and can not be definitively identified as IEM as we modeled here \cite{fischer2011}.  The PEM parameters are in reference to the Sputnik-Mamavirus-\textit{Acanthamoeba} system.

The reference values for the IEM and the PEM are shown in Tables \ref{table:cafeparams} and \ref{table:mimiparams}, respectively. Some parameters shared between the models (denoted with a $\star$) have different values because different reference sets of organisms are used for each model.  First principle derivations are shown in \ref{sec:derivation}.  Overall, these reference parameters are not well constrained based on the current literature and, as a result, we take a sampling approach within large ranges centered around our reference values to analyze the dynamics within both models (see below).


\begin{table}[h!]

\resizebox{\textwidth}{!}{
\begin{tabular}{ c | l | c | c | r}
\hline
 Symbol & Meaning & Value & units & Reference\\
  \hline
  $\star K$ & {carrying capacity} & $4.0 \times 10^6$ & $\frac{host}{ml}$ & \cite{fischer2011} \\
    $\star b$ & host birth rate & $2.7$ & $\text{day}^{-1}$ & personal communication \\
  $\star d$ & host death rate & $1.4$ & $\text{day}^{-1}$ & assumed; see \ref{sec:derivation} \\ 
    $\star m_v$ & viral decay rate & $6.3*10^{-2}$ & $\text{day}^{-1}$ & personal communication \\ 
      $\star m_p$ & virophage decay rate & $3.2*10^{-1}$ & $\text{day}^{-1}$ & personal communication\\ 
    $\star \beta_v$ & viral burst size & $130$ & $\frac{\text{viruses}}{host}$ & \cite{fischerphd} \\
    $\star \beta_{vp}=\rho_{vp} \beta_v$ & coinfecting viral burst size & $40$ & $\frac{\text{viruses}}{host}$ & personal communication\\
    $\star \beta_p=\rho_{p} \beta_v$ & virophage burst size & $1000$ & $\frac{\text{phage}}{host}$ & personal communication\\
    $\star \phi_v$ & {endocytosis of V rate} & $2.2*10^{-6}$ & $\frac{\text{ml}}{\text{virus}*\text{day}}$ & derived; see \ref{sec:derivation}\\ 
    $\phi_p$ & {endocytosis of P rate} & $1.1*10^{-5}$ & $\frac{\text{ml}}{\text{phage}*\text{day}}$ & derived; see \ref{sec:derivation}\\
              $\rho$ & fraction of infected host offspring & $0.5$ & - & personal communication \\    

\end{tabular}

}
\caption{IEM reference parameters.  The $\star$ denotes parameters shared between models.  The values of the shared parameters may differ between models as they refer to different sets of organisms.  When a value is stated for a parameter in the literature for only one system that value is used as the reference for both models as long as it is reasonable.  The parameters with \textquotedblleft$=$" show the relation between the parameters present in the model and the free parameters used for sampling.  Personal communication with M. Fischer (MPI-Heidelberg).}
\label{table:cafeparams}
\end{table}

\begin{table}[h!]
\resizebox{\textwidth}{!}{
\begin{tabular}{ c | l | c | c | r}
\hline
 Symbol & Meaning & Value & units & Reference\\
  \hline
  $\star K$ & carrying capacity & $4.0 \times 10^6$ & $\frac{host}{ml}$ & \cite{byers1980} \\
    $\star b$ & host birth rate & $1.4$ & $\text{day}^{-1}$ & \cite{byers1980} \\
  $\star d$ & host death rate & $0.70$ & $\text{day}^{-1}$ & assumed; see \ref{sec:derivation} \\ 
    $\star m_v$ & viral decay rate & $3.2*10^{-2}$ & $\text{day}^{-1}$ & \cite{campos2012} \\ 
      $\star m_p$ & virophage decay rate & $3.2*10^{-1}$ & $\text{day}^{-1}$ & personal communication\\     
    $\star \beta_v$ & viral burst size & $300$ & $\frac{\text{viral particles}}{host}$ & \cite{claverie2009} \\
    $\star \beta_{vp}=\rho_{vp} \beta_v$ & coinfecting viral burst size & $100$ & $\frac{\text{viral particles}}{host}$ & \cite{lascola2008}\\
    $\star \beta_p=\rho_{p} \beta_v$ & virophage burst size & $1500$ & $\frac{\text{viral particles}}{host}$ & approximated; see \ref{sec:derivation}\\
    $\star \phi_v$ & {absorption of V rate} & $4.3*10^{-6}$ & $\frac{\text{ml}}{\text{viral particles}*\text{day}}$ & derived; see \ref{sec:derivation}\\    
    $\phi_{vp}$ & {rate P attaches to V} & $2.2*10^{-6}$ & $\frac{\text{ml}}{\text{viral particles}*\text{day}}$ & derived; see \ref{sec:derivation}\\

        $\beta_{i}=\rho_{i} \beta_v$ & composite burst size & $[0,1]*\beta_V$ & $\frac{\text{viral particles}}{host}$ & N/A\\  
\end{tabular}}
\caption{PEM reference parameters.  The $\star$ denotes parameters shared between models.  The values of the shared parameters may differ between models as they refer to different sets of organisms.  When a value is stated for a parameter in the literature for only one system that value is used as the reference for both models as long as it is reasonable.  The parameters with \textquotedblleft$=$" show the relation between the parameters present in the model and the free parameters used for sampling.  Note \textquotedblleft N/A" denotes where information was not available.  Hence, in our statistical analysis, we sample from the full range values for $\rho_i$ that retain a reduction in the total burst size of the virus.  Personal communication with M. Fischer (MPI-Heidelberg).}
\label{table:mimiparams}
\end{table}

\subsection{Computational methods}

We utilized Latin Hypercube sampling to explore the range of dynamics possible in our models \cite{mckay1979}.  The sampling ranges were centered about the reference parameter sets in Tables \ref{table:cafeparams} and \ref{table:mimiparams} and spanned one order of magnitude above and below those values.  The only exceptions are $\rho_{vp}$, $\rho_{i}$ and $\rho$, which are bounded between 0 and 1.  From these ranges we sample $10^5$ points using the midpoints of the hypercubes.  We utilized a uniform probability distribution for parameters bounded between 0 and 1 and a log-uniform probability distribution for the other parameters for constructing our hypercubes.  We required $\rho_i+\rho_{vp} \le 1$ in the PEM to ensure that fewer viruses will be produced during virophage coinfection than without coinfection.  We sampled with respect to this constraint by uniformly, randomly sampling between $0$ and $1$ for each parameter and choosing combinations that satisfied the inequality.  We repeated the overall sampling procedure 10 times to give a total of $10^6$ sampled points for each model.

For each parameter set, equilibria of the IEM and the PEM were found using Mathematica \cite{mathematica}; script available as Supplementary File 1.  Linear stability analysis of coexistence equilibria (all state variables are positive) were also computed in Mathematica.  Linear stability analysis of boundary equilibria ($H>0$, $V>0$, other state variables are zero) were computed in MATLAB \cite{matlab}.  Similarly, simulations of the models were run using the numerical solvers ode45 or ode15s.  All simulations are available as part of Supplementary File 1 and on http://ecotheory.biology.gatech.edu/downloads.

\section{Results}
\subsection{Stable and cyclical coexistence occur given either biophysical modes of infection}
\begin{figure}[h!]

  (a)
    \includegraphics[width=0.45\textwidth]{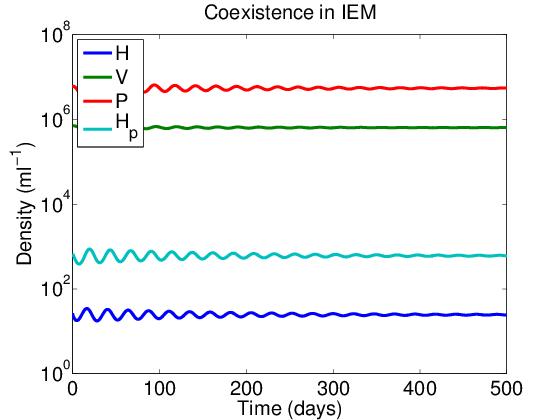}%
      (b)
    \includegraphics[width=0.45\textwidth]{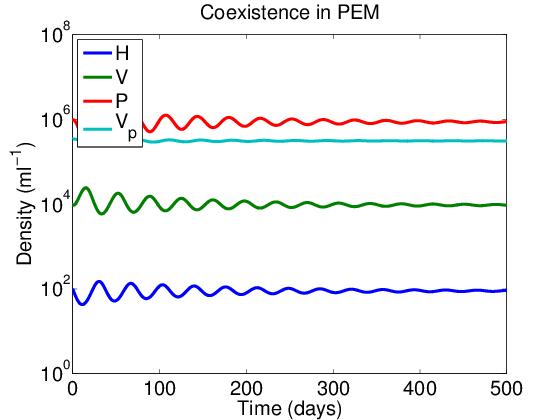}\\%
  (c)
    \includegraphics[width=0.45\textwidth]{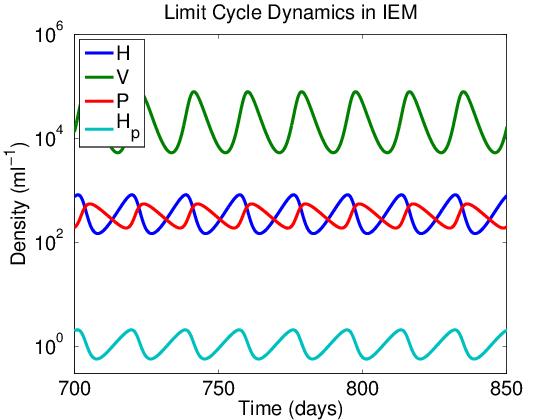}%
      (d)
    \includegraphics[width=0.45\textwidth]{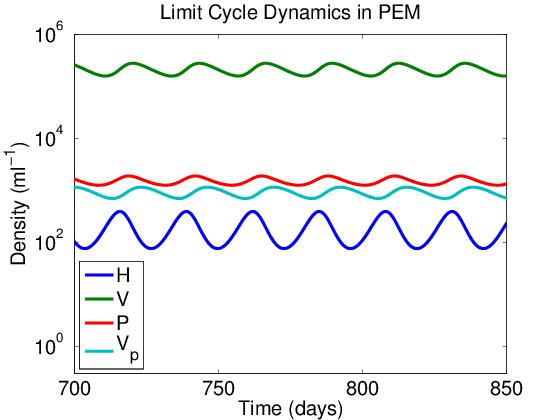}%
      \caption{Observed dynamics present in both models.  The parameter values for each figure are given in \ref{sec:parametervalues}.  The initial conditions are small random perturbations from the coexistence equilibria.  (a)  IEM stable coexistence (b) PEM stable coexistence.(c) Cyclic coexistence in the IEM. (d) Cyclic coexistence in the PEM.}\label{fig:coexdynamics} 
\end{figure}

Coexistence equilibria arise in systems \eqref{eqn:IEM} and \eqref{eqn:PEM}, when the right hand sides of those systems are zero for positive densities of the host, virophage and virus.  In both systems, coexistence equilibria can be stable or unstable.  We observe for some cases where the coexistence equilibria are unstable that the species exhibit cyclic dynamics.  Examples of stable coexistence for both models are shown in Figures \ref{fig:coexdynamics}(a,b).  Examples of cycle coexistence are shown in Figures \ref{fig:coexdynamics}(c,d).

\begin{figure}[h!]
      (a)
    \includegraphics[width=0.45\textwidth]{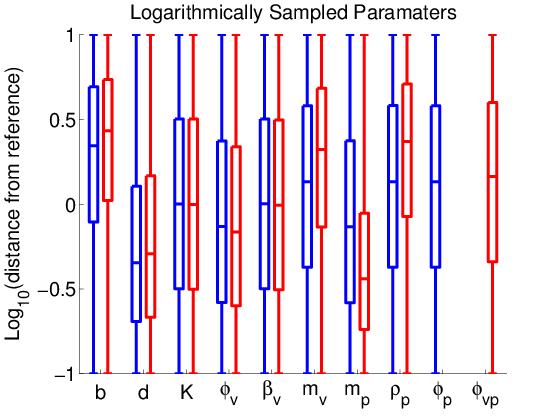}%
          (b)
    \includegraphics[width=0.45\textwidth]{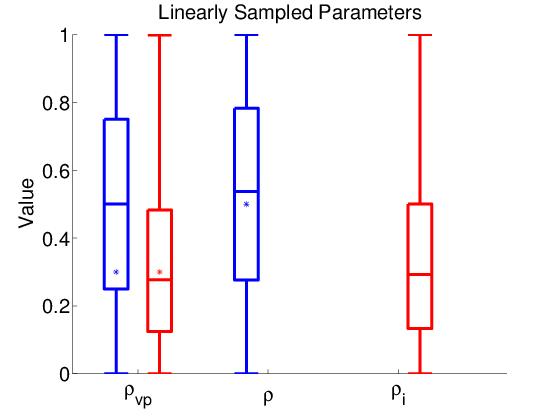}%
      \caption{Marginal distributions of the parameters for the cases when stable coexistence occurs.  Blue (red respectively) boxplots correspond to the IEM (PEM respectively).  The median of the distributions are the center lines with the edges corresponding to the 25th and 75th percentile, and the tails extend to the minimum and maximum of the distributions.  Shared parameters feature two box-plots for each label and unique parameters feature one box plot.  (a) log base 10 distribution relative to the reference parameter sets in Tables \ref{table:mimiparams} and \ref{table:cafeparams} for logarithmically sampled parameters. (b) Distributions of linearly sampled parameter values.  The reference parameter values are denoted by an asterisk (*).  No reference parameter value is used for $\rho_i$.}\label{fig:coexparams} 
\end{figure}

Next we considered the statistical nature of when stable coexistence occurs.  We only considered coexistence points where virophage and infected class abundances were each greater than $10^{-7} \text{ml}^{-1}$.  Boxplots of the parameter distributions in Figure \ref{fig:coexparams} show ranges for each parameter value that allow for stable coexistence.  For the parameters sampled in log-space the box plots represent the base 10 logarithm of the marginal distributions in terms of the distance from the reference parameter set.  For example, consider the box plot for the marginal distribution of the birth rate of the host, $b$, in the PEM, shown in red (Figure \ref{fig:coexparams}a).  The median of the distribution is nearly $0.5$, meaning almost half of the sampled parameter sets for which coexistence occurs have a birth rate over half an order of magnitude ($\sim$3 times) larger than the reference value.  Further, the 25th and 75th percentiles of the distributions (edges of boxes in Figure 3a) are above zero, implying that 75\% of the sampled parameter sets for which coexistence occurs have a birth rate that is greater than the reference value.  For the linearly sampled parameter values (Figure \ref{fig:coexparams}b), the box plots represent the marginal distributions of the parameters.  The reference values for each parameter are marked with an asterisk (*).  Overall, the reference values are contained within the middle 50th percentile for most of the parameters; exceptions include $b$ and $m_p$ in the PEM.  Coexistence tends to occur when parameters are beneficial to the host and virophage (e.g., high b, high $\phi_p$/$\phi_{vp}$, high $\rho_p$, low d, low $m_p$) and parameters specific to viruses are detrimental (e.g., low $\phi_{v}$, high $m_v$), when compared to baseline parameter values.

\subsection{Virophage presence increases host abundance and decreases viral abundance}

\begin{figure}[h!]

      (a)
    \includegraphics[width=0.45\textwidth]{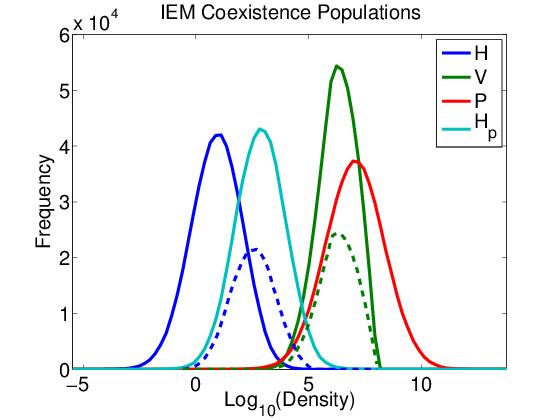}%
      (b)
    \includegraphics[width=0.45\textwidth]{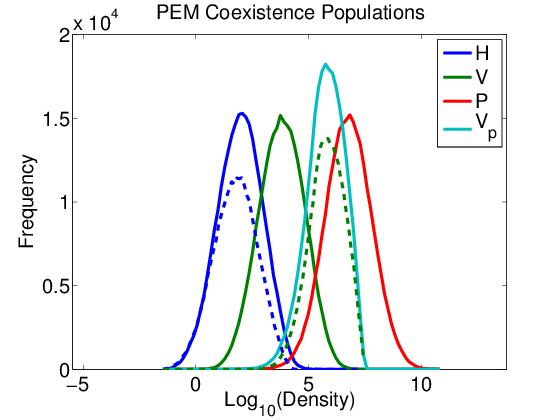}%
      \caption{Histograms of stable coexistence populations in each model.  Dashed lines are the histograms for the respective boundary equilibrium with host and virus alone.  Units for the transformed densities on the x-axis are $\text{ml}^{-1}$.  (a)  IEM (b) PEM.}\label{fig:histpop} 
\end{figure}

Histograms of population densities for stable equilibrium points are shown in Figure \ref{fig:histpop}.  Virophage tend to be the most abundant entity for both models.  Additionally, the infected classes (solid cyan lines), which represent the virophage associated with host or virus, tend to be larger than the respective uninfected class for both models (blue line in Figure \ref{fig:histpop}a, green line in Figure \ref{fig:histpop}b).  These results suggest that virophage will be the most abundant entity in field measurement data.  However, counterexamples exist where viruses are more abundant than virophage at equilibrium.  As a result, field measurements may be useful in determining the covariation between parameters.  For example, abundance data could suggest elimination of  parameter sets that feature incorrect rank abundance of the populations.  The dashed histograms in Figure \ref{fig:histpop} are the population densities for the boundary equilibria.  The presence of the virophage causes the host and virus histograms to shift; however, the effect on the total amount of hosts and viruses is not clear and is addressed below.

To address the effect of virophages on the hosts we define the total host abundance as the abundance of host genomes present.  For the IEM, a member of the infected class contains a host genome and thus we define the total host abundance as $H^*_{\text{total}}=H^*+H^*_p$, where the (*) denotes equilibrium densities.  For the PEM, the hosts are the only modeled variable that involves a host genome and, hence, $H^*_{\text{total}}=H^*$.  Similarly, to address the effect of virophages on the viruses we define the total viral abundance as the abundance of viral genomes present.  For the PEM, a member of the infected class contains a viral genome and we define the total viral abundance as $V^*_{\text{total}}=V^*+V^*_p$.  For the IEM, the viruses are the only modeled variable that involves a viral genome and, hence, $V^*_{\text{total}}=V^*$.

\begin{figure}[h!]

  (a)
    \includegraphics[width=0.45\textwidth]{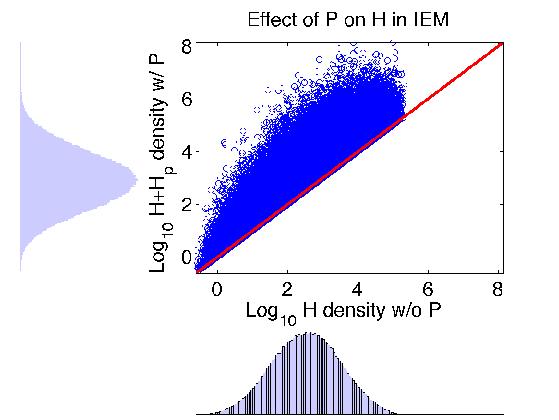}%
      (b)
    \includegraphics[width=0.45\textwidth]{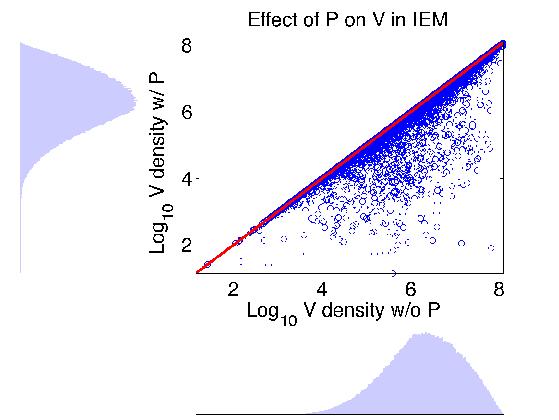}\\
      (c)
    \includegraphics[width=0.45\textwidth]{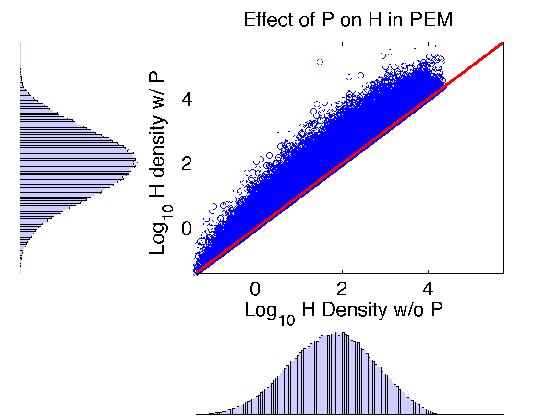}%
      (d)
    \includegraphics[width=0.45\textwidth]{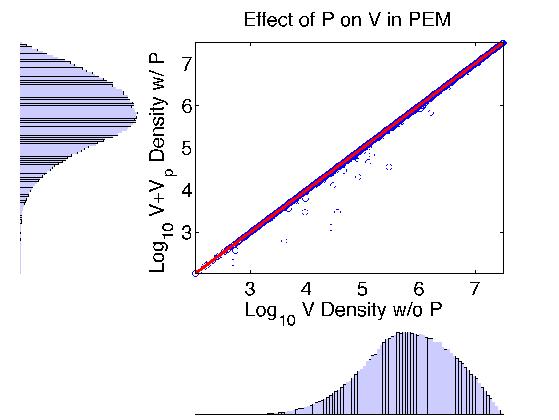}%
      \caption{Comparing the marginal distributions of the species' genome abundances at the coexistence equilibria and the boundary equilibria where only the host and virus are present.  Infected classes are combined with uninfected classes for comparison of total genome abundances (e.g., $V_{\text{total}} = V+V_p$ in the PEM).  Units for the transformed densities on each axis are $\text{ml}^{-1}$.  Effect of virophage on (a) host genome abundance and (b) viral genome abundance in the IEM. Effect of virophage on (c) host genome abundance and (d) viral genome abundance in the PEM.}\label{fig:hvcompare}
\end{figure}

We compare the equilibrium total abundances of the host and viral populations in the presence and absence of the virophage in Figure \ref{fig:hvcompare}.  Note we only consider parameter sets where the coexistence equilibrium points are stable.  In Figure \ref{fig:hvcompare}, the red lines are the 1-1 line, where virophage has no effect on the host and virus abundances.  Since all points in figure \ref{fig:hvcompare}(a,c) lie above the 1-1 line, the virophage increases the equilibrium density of the host.  Since all points in figure \ref{fig:hvcompare}(b,d) lie below the 1-1 line, the virophage reduces the abundance of the virus.  We note that the relative increases in host abundance and relative decreases in virus abundance tend to be greater in magnitude for the IEM.  In total, irrespective of infection mode, the effect of the virophage on equilibrium density can be summarized as the virus of a host's virus is the host's \textquotedblleft friend."

To see why virophage always reduce the total viral density and increase the total host density at equilibrium, consider the respective dynamics.  We define an average burst size parameter: $\bar{\beta_v} = \frac{\beta_v H + \beta_{vp} H_p}{H+H_p}$ for the IEM, and $\bar{\beta_v} = \frac{\beta_v V + (\beta_{vp}+\beta_i) V_p}{V+V_p}$ for the PEM.  Note that $\bar\beta_v \le \beta_v$ in both models.  The dynamics for the total host and total virus densities simplify to a standard competitive Lotka-Volterra system where the virus is a predator of the host (derived in \ref{sec:lumpedmodel}):

\begin{eqnarray}\label{eqn:lumped}\begin{split}
\dot H_{total} &= H_{total} \left[b-d\left(1+\frac{H_{total}}{K}\right)\right] - \phi_v V_{total} H_{total} \\
\dot V_{total} &= \bar{\beta_{v}} \phi_{v} H_{total} V_{total} - m_v V_{total}.&
\end{split}\end{eqnarray}
The parameters of system \eqref{eqn:lumped} are the same as in models \eqref{eqn:IEM} and \eqref{eqn:PEM}.  Solving for the equilibrium populations (denoted with $*$) and utilizing the bound on $\bar \beta_v$, we obtain:

\begin{eqnarray}
H^*_{total} &=& \frac{m_v}{\phi_v \bar{\beta^*_v}}\ge \frac{m_v}{\phi_v \beta_v}=H^*_b\\
V^*_{total} &=& \frac{1}{\phi_v} \left[b-d\left(1+\frac{m_v}{\phi_v \bar{\beta^*_v}K}\right)\right] \nonumber \\ 
		&\le& \frac{1}{\phi_v} \left[b-d\left(1+\frac{m_v}{\phi_v \beta_v K}\right)\right]=V^*_b,
\end{eqnarray}
where the subscript $b$ refers to the boundary equilibrium with hosts and viruses only.  The burst size of the genome level model at equilibrium is represented by $\bar{\beta^*_v}$.  Thus, virophage coinfection causes a reduction in burst size, which in turn, increases total host abundance and decreases total viral abundance.  This effect occurs so long as the virophage has a deleterious effect on the burst size of the virus, i.e., $\beta_{vp}<\beta_v$ for the IEM and $\beta_{vp}+\beta_i<\beta_v$ for the PEM (see  \ref{sec:lumpedmodel}).

\subsection{Bistability in PEM} 
\begin{figure}
\begin{minipage}{0.5\textwidth}
(a)
    \includegraphics[width=\textwidth]{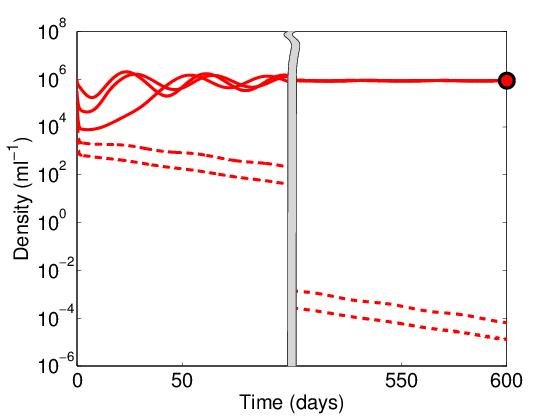}\\
      (b)
    \includegraphics[width=\textwidth]{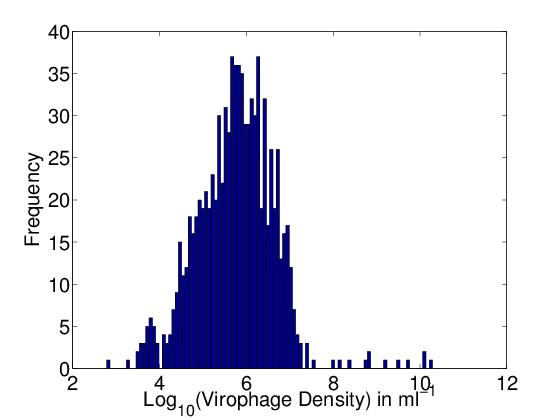}\\
          (c)
    \includegraphics[width=\textwidth]{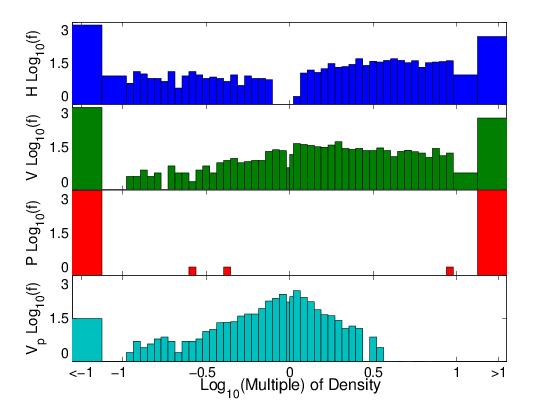}
\end{minipage}%
\hspace{0.5cm}
\begin{minipage}{0.5\textwidth}
\caption{Bistability in the PEM.  (a)  Virophage dynamics given different initial conditions.  The initial conditions are that the host and virus are at the boundary equilibrium ($H_0 \approx 23.0$ and $V_0 \approx 3.20*10^{5}$) and the virophage varies as $P_0 = 10^{m}$ where $m \in \{4,4.5,5,5.5,6\}$.  Dynamics that lead to coexistence (virophage crashing) are plotted with solid (dashed) lines.  The red circle shows the theoretical virophage coexistence population.  (b)  Histogram of the approximate distance from the boundary equilibrium to the edge of its basin of attraction in the direction of virophage density.  We present the data for 907 random parameter sets that yielded convergence. (c)  Histograms of the logarithm of frequencies of approximate locations of the basin of attraction boundaries of the coexistence equilibrium with respect to each population.  Histograms of the approximate distance from the coexistence equilibria to the edge of its basin of attraction in the direction of the different state variables. Data represents 1000 random parameter sets that allow bistability and where the basin of attraction boundaries lie between 0.1 and 10 times the respective coexistence population densities. Distances were computed in terms of a multiple of the respective coexistence equilibrium population density. The  abscissa is the logarithm of the multiple of the respective coexistence equilibrium population density.  Values below (above) 0 mean that the boundary is located at density values less (greater) than the coexistence equilibrium density. The ordinate is the logarithm of the number of parameter sets.  Points that lie outside the range of 0.1 and 10 times the equilibrium density are binned together and denoted as $<-1$ and $>1$ respectively.
\label{fig:bistability}}
\end{minipage}
\end{figure}

In our numerical simulations we did not find parameter values for which bistability arises in the IEM.  In contrast, approximately $1\%$ of the parameter sets in the PEM yielded bistability.  Bistability arises when two equilibria are locally stable, and results in asymptotic dynamics dependent on initial conditions.  An example of this bistability between two ranges of time is shown in Figure \ref{fig:bistability}a.  When smaller amounts of virophage are added, the virophage are unable to invade (two dashed curves), whereas when amounts greater than a certain amount (here $\sim 10^{4.5} \text{ml}^{-1}$) are added the virophage are able to invade (three solid curves).  In most cases where bistability was observed, the boundary and the coexistence equilibria were both locally stable fixed points.  In these cases, there also existed a second coexistence equilibrium that was saddle point (i.e., semistable).  In a few other cases, bistability arose when the boundary equilibrium was stable and both of the coexistence equilibria were locally unstable.  In these few cases, there was cyclic coexistence between the host, virophage and virus.  We note that this case was difficult to find numerically.  Hence, in the following we focus on parameter sets where one coexistence equilibrium and the boundary equilibrium are locally stable.


For parameter sets with bistability there exists a basin of attraction for the coexistence point in phase space.  We interpret the size of this basin of attraction as a proxy for the robustness of coexistence to environmental perturbations.  Hence, we identified the boundary of this basin of attraction along the axes of phase space from the boundary equilibrium (Figure \ref{fig:bistability}b) and from the coexistence point (Figure \ref{fig:bistability}c).  The boundary of the basin of attraction has a different interpretation in each case, as discussed individually below.

Figure \ref{fig:bistability}b shows a histogram of the boundary of the basin of attraction along the virophage axis from the boundary equilibrium.  An interpretation of this boundary in phase space is the minimum amount of virophage required to invade a system at equilibrium with hosts and viruses alone.  These results were obtained by randomly sampling 1000 parameter sets where bistability is expected to occur based on the linear stability analysis.  We repeatedly simulated the dynamics with the boundary equilibrium for hosts and viruses and varying amounts of virophage as the initial condition.  We performed a bisection method in log space for initial amounts of virophage with a range of [$10^{-4}$, $10^4$] times the virophage population at the coexistence equilibrium.  Out of the 1000 samples, for 93 simulations either the basin of attraction was outside the range of our bisection method or the dynamics did not converge to virophage invasion or crashing within a specified time.  We did not include these parameter sets in our histogram giving a total of 907 parameter sets in the data.  For the remaining parameter sets, the average of the minimal amount of virophage added that led to coexistence and maximum amount of added virophage that led to virophage extinction are the values in the histogram.  These values are accurate within .005 in the log space range of the prefactor as mentioned above.  Overall, this figure illustrates that a non-negligible amount of virophage must be introduced in order for coexistence to occur.

Figure \ref{fig:bistability}c shows histograms of the boundary of the basin of attraction along the phase space axes from the coexistence equilibrium for our 1000 samples.  The values of the x-axis are relative to the respective coexistence equilibrium population.  An interpretation of this boundary in phase space is a bound on the amount of each respective population that can be added or removed without causing the virophage to crash.  These values were obtained using a bisection method similar to the one previously described.  One difference is a smaller range was used ([$10^{-1}$, $10^1$] times the respective coexistence population).  This range spans from reducing the respective population to 10\% of its coexistence value to increasing the respective population to 10 times its coexistence value.  We bin together the parameter sets for which the boundary existed outside our range and include this in our histograms.

Figure \ref{fig:bistability}c shows that coexistence is differentially robust to perturbations of the different population densities.  For example, coexistence is very robust to removal of virophage as nearly all of the parameter sets maintained coexistence within the entire range of perturbations.  In comparison, the peak of the $V_p$ histograms  are close to $0$ and almost entirely contained within our range.  This suggests virophage coexistence, when it occurs ina region of parameter space corresponding to bistability, is highly sensitive to addition or removal of virus-virophage composites.


\subsection{Sampled coexistence points tend to be attracting} 

We now investigate the frequency of stable coexistence in our parameter sets.  We used linear (local) stability of the coexistence equilibria and the boundary equilibria (equilibria with host and virus alone) to identify if the equilibria were stable (\textquotedblleft S") or unstable (\textquotedblleft U").   A statistical enumeration of the linear stability of the equilibria is shown in Table \ref{table:stats}.  Table \ref{table:stats}a corresponds to the IEM.  The columns of Table \ref{table:stats}a define the stability of the boundary equilibrium of the IEM and the rows of Table \ref{table:stats}a define the stability of the coexistence point of the IEM.  Table \ref{table:stats}b corresponds to the PEM.  The columns of Table \ref{table:stats}b define the stability of the boundary equilibrium of the PEM.  Since multiple coexistence equilibria can arise in the PEM, the rows of Table \ref{table:stats}b are divided into cases where there is one coexistence equilibrium (\textquotedblleft single coexist") or two coexistence equilibria (\textquotedblleft multi coexist").

We make a few points about Table \ref{table:stats}.  First, stable coexistence occurs when at least one of the coexistence equilibria is stable.  When all of the coexistence equilibria are unstable, then cyclic coexistence, aperiodic coexistence, or extinction of the virophage are possible outcomes.  Second, out of the $10^6$ parameter sets, coexistence equilibria (either stable or unstable) are observed for approximately half of the parameter sets for the IEM and approximately a quarter of the parameter sets for the PEM.  Thus, based on these totals, a larger portion of the parameter space allows for coexistence in the IEM versus the PEM.

Third, since the \textquotedblleft SS" row in Table \ref{table:stats}b sums to zero, bistability between coexistence points was not observed in the PEM.  Thus, for our parameter ranges, bistability only occurs when both a boundary equilibrium and a coexistence equilibrium are locally stable.  Since bistability was not observed in the IEM, we interpret these results to suggest that the coexistence of virophage is more robust to perturbations in the IEM versus the PEM. Finally, cyclic coexistence can only occur when all coexistence equilibria are unstable.  Stable coexistence equilibria occur at a higher frequency than unstable equilibria in the IEM (compare \textquotedblleft S" row and \textquotedblleft U" row in Table \ref{table:stats}a).  Similarly, in the PEM, it is more frequent that at least one coexistence equilibrium is stable than all coexistence equilibria being unstable (compare \textquotedblleft S" and \textquotedblleft SU/US" rows to \textquotedblleft U" and \textquotedblleft UU" rows in Table \ref{table:stats}b).  Thus, the dynamics observed in our models suggest that if cycles are observed in experimental population dynamics, then either the parameters of the systems are finely tuned or the cycles are forced by sources outside of our modeling framework (e.g., predation on the host species).


\begin{table}
(a)
\resizebox{.45\textwidth}{!}{
\begin{tabular}{| c | c | c c |}
\hline
IEM & & \multicolumn{2}{ |c| }{Boundary} \\
\hline
   &  & S & U \\
  \hline
   \multirow{2}{*}{Coexist} & S & 0 & 498821 \\
   & U & 0 & 34 \\  
  \hline
  & Total & 0 & 498855 \\
  \hline
   \end{tabular}}
(b)
\resizebox{.45\textwidth}{!}{
\begin{tabular}{| c | c | c c |}
\hline
\multicolumn{2}{ |c| }{PEM}  & \multicolumn{2}{ |c| }{Boundary} \\
\hline
 & & S & U  \\
  \hline
  \multirow{2}{*}{Single Coexist} & S & 0 & 243324  \\
  & U & 0 & 9131   \\
  \hline
  \multirow{3}{*}{Multi Coexist} & SS & 0 & 0   \\
  & SU/US & 11342 & 0   \\
  & UU & 77 & 0   \\
    \hline
   & Total & 11419 & 252455  \\
    \hline
   \end{tabular}
   }
   \caption{Frequency of stable coexistence in the IEM and PEM models.  Each table lists the number of parameter points satisfying different combinations of linear stabilities for the coexistence points and the boundary equilibria for (a) IEM and (b) PEM.  The total number of sampled parameter sets for each model is $10^6$.  \textquotedblleft S" refers to stable and \textquotedblleft U" refers to unstable. (b) In the PEM, parameter sets can have one(\textquotedblleft Single Coexist") or multiple(\textquotedblleft Multi Coexist")  equilibrium points.  Rows with multiple letters denote the stabilities of the two coexistence points.}
\label{table:stats}
   \end{table}

\section{Discussion}

The study of virophage interactions with viruses is at its infancy.  Nonetheless, multiple independent discoveries have been made of virophage populations persisting with viruses and their eukaryotic hosts in a diverse range of environments from cooling towers to the open oceans.  These discoveries motivated our central aim to develop biophysically-motivated models of the interactions among virophage, viruses and their eukaryotic hosts and to understand what effects virophage have on the dynamics of these populations. The models we proposed correspond to two distinct cases: (i) where virophage attach to viruses and then the composite infects host cells (the PEM); and (ii)  where viruses and virophage independently infect host cells (the IEM).  Coexistence amongst all populations is possible in both models when analyzed over plausible ranges of parameter space.  In addition,  both models allowed for stable and cyclical asymptotic dynamics.  Importantly, we demonstrated both analytically and numerically that so long as virophage negatively affect virus burst size then virophage will act as the \textquotedblleft friend" of their hosts, i.e., increasing host abundance and decreasing virus abundance, irrespective of infection mechanism. 

These results add an ecological layer to prior observations of the cellular level effects of virophage on viruses and eukaryotic hosts.  They also suggest testable hypotheses for evaluating the ecological effects of the presence of virophage within communities (e.g., when virophage enter a new environment, they should drive viral populations down resulting in an increase of host populations).   The models may also be used to help distinguish between the biophysical mode of infection where virophage are present.  For example, bistability was observed only in the PEM.  An experimentalist may test for bistability by first obtaining coexistence between virus and host within a chemostat and then observing both extinction and coexistence of virophage after introducing different concentrations of virophage.  If bistability is observed with virophage that follow the PEM then stochastic fluctuations may be more likely to lead to virophage extinction in comparison to systems where the virophage follow the IEM.

An alternative approach to distinguishing between infection modes arises from analyzing phase lags in those instances where cyclical dynamics are observed.  A similar approach has been proposed to distinguish between indirect and direct transmission in the spread of infectious pathogens within traditional epidemiological SIR-type models~\cite{cortez2013}.    Here, we have limited preliminary evidence to suggest a similar approach may also be of use.   We observed that the virus population cycles preceded the virophage population cycles in the IEM, whereas the virophage cycles preceded the virus cycles in the PEM.  By ``precede'', we mean that the population maximum (and minimum) of one type appears immediately before the population maximum (and minimum) of the other type.   Hence, measurements the densities and orderings of peaks for the virus and virophage could help distinguish between infection modes.  However, our analysis involves a small number of examples; initial exploration is presented in \ref{sec:cycles} and warrants follow-up study.

Although the models developed here were constructed with virophage in mind, they may be useful in modeling the ecological effect of satellite viruses or other defective interfering particles.  In fact, a previous model of defective interfering particles shares a similar form to our independent entry model; however, it differs in construction and analysis whereby infections were treated as stage structured and simulations assumed an \textit{in vitro} setting where viruses were repeatedly introduced through passages \cite{kirkwood1994}.

These models may also be relevant to other organisms since virophage function as a special case of hyperparasitism.  In our case the virophage functions as the hyperparasitoid.  Previous models of hyperparasitism have been limited to two classes of models: epidemiological type models where the parasite and hyperparasite spread through direct transmission of the hosts \cite{morozov2007, holt1998} and population models based on difference equations \cite{may1981, beddington1977}.

In moving forward, it is important to note a secondary contribution of this study: the establishment of parameter baselines applicable to distinct biophysical modes of virophage-virus-host interactions.  In sampling parameter space we assumed parameter distributions independent from each other.  In reality, both the range of parameter values and their covariation are likely to be more constrained as a result of trade-offs, biophysical limits, and other effects.  We suggest the need for further empirical studies to refine both the qualitative and quantitative nature of these interactions.  Such refinement is likely to provide further evidence to establish 
when environments are likely to support a virophage population in the first place, 
identify ecological effects common to both modes, 
and identify which of our proposed means for distinguishing the mode of coinfection from population level data are useful.  Additionally, given better constraints on the potential range of parameter values, we will also be able to extend the current model to ask evolutionary questions, e.g., how virophage interactions may evolve in distinct ecological contexts. 
In doing so, extending the current framework to a spatially explicit context is 
likely to be of use as spatial models stabilize viral-host systems and can yield alternative conclusions to evolutionary questions \cite{heilmann2010}.  Extension to a spatial model seems particularly relevant for virophage given the requirement of host coinfection
within a large, complex population
(whether infecting together or independently) in order for virophage to reproduce.

\section{Acknowledgments}

The authors thank Matthias Fischer for helpful discussions leading to the establishment of baseline values for key parameters in this model, as well as additional feedback on the manuscript.  The authors thank C. Flores and L. Jover for feedback on the manuscript.  This work was supported, in part, by a grant from the James S McDonnell Foundation.  MHC was supported by the National Science Foundation under Award No. DMS-1204401.   JSW holds a Career Award at the Scientific Interface from the Burroughs Wellcome Fund.

\bibliographystyle{plain}
\bibliography{VoVbib2}

\begin{thebibliography}{10}

\bibitem{beddington1977}
J.R. Beddington and P.S. Hammond.
\newblock {On the dynamics of host-parasite-hyperparasite interactions}.
\newblock {\em {Journal of Animal Ecology}}, {46}({3}):{811--821}, {1977}.

\bibitem{beretta2001}
E~Beretta and Y~Kuang.
\newblock {Modeling and analysis of a marine bacteriophage infection with
  latency period}.
\newblock {\em {Nonlinear Analysis-Real World Applications}},
  {2}({1}):{35--74}, {Mar} {2001}.

\bibitem{bergpurcell1977}
H.C. Berg and E.M. Purcell.
\newblock Physics of chemoreception.
\newblock {\em Biophysical Journal}, 20(2):193--219, 1977.

\bibitem{boyer2011}
Michael Boyer, Said Azza, Lina Barrassi, Thomas Klose, Angelique Campocasso,
  Isabelle Pagnier, Ghislain Fournous, Audrey Borg, Catherine Robert, Xinzheng
  Zhang, Christelle Desnues, Bernard Henrissat, Michael~G. Rossmann, Bernard
  La~Scola, and Didier Raoult.
\newblock {Mimivirus shows dramatic genome reduction after intraamoebal
  culture}.
\newblock {\em {Proceedings of the National Academy of Sciences of the United
  States of America}}, {108}({25}):{10296--10301}, {JUN 21} {2011}.

\bibitem{byers1980}
TJ~Byers, RA~Akins, BJ~Maynard, RA~Lefken, and SM~Martin.
\newblock {Rapid growth of \textit{Acanthamoeba} in defined media -- induction
  of encystment by glucose-acetate starvation}.
\newblock {\em {Journal of Protozoology}}, {27}({2}):{216--219}, {1980}.

\bibitem{campos2012}
Rafael~Kroon Campos, Ketyllen~Reis Andrade, Paulo~Cesar Peregrino~Ferreira,
  Claudio~Antonio Bonjardim, Bernard La~Scola, Erna~Geessien Kroon, and
  Jonatas~Santos Abrahao.
\newblock {Virucidal activity of chemical biocides against mimivirus, a
  putative pneumonia agent}.
\newblock {\em {Journal of Clinical Virology}}, {55}({4}):{323--328}, {DEC}
  {2012}.

\bibitem{claverie2009}
J.~M. Claverie, C.~Abergel, and H.~Ogata.
\newblock {Mimivirus}.
\newblock In {VanEtten, JL}, editor, {\em {Lesser Known Large DSDNA Viruses}},
  volume {328} of {\em {Current Topics in Microbiology and Immunology}}, pages
  {89--121}. {2009}.

\bibitem{cohen2011}
Gaelle Cohen, Louis Hoffart, Bernard La~Scola, Didier Raoult, and Michel
  Drancourt.
\newblock {Ameba-associated Keratitis, France}.
\newblock {\em {Emerging Infectious Diseases}}, {17}({7}):{1306--1308}, {JUL}
  {2011}.

\bibitem{cortez2013}
Michael~H. Cortez and Joshua~S. Weitz.
\newblock Distinguishing between indirect and direct modes of transmission
  using epidemiological time series.
\newblock {\em American Naturalist}, 181(2):E43--E52, 2013.

\bibitem{desnues2010}
C.~Desnues and D.~Raoult.
\newblock Inside the lifestyle of the virophage.
\newblock {\em Intervirology}, 53(5):293--303, 2010.

\bibitem{desnues2012}
Christelle Desnues, Bernard La~Scola, Natalya Yutin, Ghislain Fournous,
  Catherine Robert, Said Azza, Priscilla Jardot, Sonia Monteil, Angelique
  Campocasso, Eugene~V. Koonin, and Didier Raoult.
\newblock {Provirophages and transpovirons as the diverse mobilome of giant
  viruses}.
\newblock {\em {Proceedings of the National Academy of Sciences of the United
  States of America}}, {109}({44}):{18078--18083}, {OCT 30} {2012}.

\bibitem{desnues2012b}
Christelle Desnues and Didier Raoult.
\newblock {Virophages question the existence of satellites}.
\newblock {\em {Nature Reviews Microbiology}}, {10}({3}):{234}, {MAR} {2012}.

\bibitem{fischer2010}
Matthias~G. Fischer, Michael~J. Allen, William~H. Wilson, and Curtis~A. Suttle.
\newblock {Giant virus with a remarkable complement of genes infects marine
  zooplankton}.
\newblock {\em {Proceedings of the National Academy of Sciences of the United
  States of America}}, {107}({45}):{19508--19513}, {NOV 9} {2010}.

\bibitem{fischer2011}
Matthias~G. Fischer and Curtis~A. Suttle.
\newblock {A virophage at the origin of large DNA transposons}.
\newblock {\em {Science}}, {332}({6026}):{231--234}, {APR 8} {2011}.

\bibitem{fischerphd}
Matthias~Gunther Fischer.
\newblock {\em Genetic and ultrastructural characterization of Cafeteria
  roenbergensis virus and its virophage Mavirus}.
\newblock PhD thesis, University of British Columbia, 2011.

\bibitem{gaia2013}
Morgan Gaia, Isabelle Pagnier, Angelique Campocasso, Ghislain Fournous, Didier
  Raoult, and Bernard La~Scola.
\newblock {Broad spectrum of mimiviridae virophage Allows Its Isolation Using a
  Mimivirus Reporter}.
\newblock {\em {PloS One}}, {8}({4}), {APR 15} {2013}.

\bibitem{heilmann2010}
S.~Heilmann, K.~Sneppen, and S.~Krishna.
\newblock Sustainability of virulence in a {Phage-Bacterial} ecosystem.
\newblock {\em Journal of Virology}, 84:3016--3022, 2010.

\bibitem{holt1998}
RD~Holt and ME~Hochberg.
\newblock {The coexistence of competing parasites. Part II - Hyperparasitism
  and food chain dynamics}.
\newblock {\em {Journal of Theoretical Biology}}, {193}({3}):{485--495}, {AUG
  7} {1998}.

\bibitem{seawater}
George William~Clarkson Kaye and Thomas~Howell Laby.
\newblock {\em Tables of Physical \& Chemical Constants}.
\newblock Longman, 16th edition, 1995.

\bibitem{kirkwood1994}
TBL Kirkwood and CRM Bangham.
\newblock {Cycles, chaos, and evolution in virus cultures - a model of
  defective interfering particles}.
\newblock {\em {Proceedings of the National Academy of Sciences of the United
  States of America}}, {91}({18}):{8685--8689}, {AUG 30} {1994}.

\bibitem{lascola2008}
Bernard La~Scola, Christelle Desnues, Isabelle Pagnier, Catherine Robert, Lina
  Barrassi, Ghislain Fournous, Michele Merchat, Marie Suzan-Monti, Patrick
  Forterre, Eugene Koonin, and Didier Raoult.
\newblock {The virophage as a unique parasite of the giant mimivirus}.
\newblock {\em {Nature}}, {455}({7209}):{100--104}, {SEP 4} {2008}.

\bibitem{levin1977}
BR~Levin, FM~Stewart, and L~Chao.
\newblock {Resource-limited growth, competition, and predation - A model and
  experimental studies with bacteria and bacteriophage}.
\newblock {\em {American Naturalist}}, {111}({977}):{3--24}, {1977}.

\bibitem{massana2007}
Ramon Massana, Javier del Campo, Christian Dinter, and Ruben Sommaruga.
\newblock {Crash of a population of the marine heterotrophic flagellate
  Cafeteria roenbergensis by viral infection}.
\newblock {\em {Environmental Microbiology}}, {9}({11}):{2660--2669}, {NOV}
  {2007}.

\bibitem{mathematica}
Mathematica.
\newblock {\em version 9.0.1.0}.
\newblock Wolfram Research, Inc., Champaign, Illinois, 2013.

\bibitem{matlab}
MATLAB.
\newblock {\em version 8.0.0.783 (R2012b)}.
\newblock The MathWorks Inc., Natick, Massachusetts, 2012.

\bibitem{may1981}
RM~May and MP~Hassell.
\newblock {The dynamics of multiparasitoid-host interactions}.
\newblock {\em {American Naturalist}}, {117}({3}):{234--261}, {1981}.

\bibitem{mckay1979}
MD~Mckay, RJ~Beckman, and WJ~Conover.
\newblock {A comparison of three methods for selecting values of input
  variables in the analysis of output from a computer code}.
\newblock {\em {Technometrics}}, {21}({2}):{239--245}, {1979}.

\bibitem{morozov2007}
Andrew~Yu. Morozova, Cecile Robin, and Alain Franc.
\newblock {A simple model for the dynamics of a host-parasite-hyperparasite
  interaction}.
\newblock {\em {Journal of Theoretical Biology}}, {249}({2}):{246--253}, {NOV
  21} {2007}.

\bibitem{mutsafi2010}
Yael Mutsafi, Nathan Zauberman, Ilana Sabanay, and Abraham Minsky.
\newblock {Vaccinia-like cytoplasmic replication of the giant Mimivirus}.
\newblock {\em {Proceedings of the National Academy of Sciences of the United
  States of America}}, {107}({13}):{5978--5982}, {MAR 30} {2010}.

\bibitem{ogataclaverie2008}
Hiroyuki Ogata and Jean-Michel Claverie.
\newblock {Microbiology - How to infect a mimivirus}.
\newblock {\em {Science}}, {321}({5894}):{1305--1306}, {SEP 5} {2008}.

\bibitem{sun2010}
Siyang Sun, Bernard La~Scola, Valorie~D. Bowman, Christopher~M. Ryan, Julian~P.
  Whitelegge, Didier Raoult, and Michael~G. Rossmann.
\newblock {Structural Studies of the Sputnik Virophage}.
\newblock {\em {Journal of Virology}}, {84}({2}):{894--897}, {JAN 15} {2010}.

\bibitem{vanetten2010}
James~L. Van~Etten, Leslie~C. Lane, and David~D. Dunigan.
\newblock Dna viruses: the really big ones (giruses).
\newblock {\em Annual Review of Microbiology}, 64:83--99, 2010.

\bibitem{wodarz2013}
Dominik Wodarz.
\newblock {Evolutionary dynamics of giant viruses and their virophages}.
\newblock {\em {Ecology and Evolution}}, {3}({7}):{2103--2115}, {JUL} {2013}.

\bibitem{xiao2005}
CA~Xiao, PR~Chipman, AJ~Battisti, VD~Bowman, P~Renesto, D~Raoult, and
  MG~Rossmann.
\newblock {Cryo-electron microscopy of the giant mimivirus}.
\newblock {\em {Journal of Molecular Biology}}, {353}({3}):{493--496}, {OCT 28}
  {2005}.

\bibitem{xiao2009}
Chuan Xiao, Yurii~G. Kuznetsov, Siyang Sun, Susan~L. Hafenstein, Victor~A.
  Kostyuchenko, Paul~R. Chipman, Marie Suzan-Monti, Didier Raoult, Alexander
  McPherson, and Michael~G. Rossmann.
\newblock {Structural studies of the giant mimivirus}.
\newblock {\em {PLoS Biology}}, {7}({4}):{958--966}, {APR} {2009}.

\bibitem{yau2011}
Sheree Yau, Federico~M. Lauro, Matthew~Z. DeMaere, Mark~V. Brown, Torsten
  Thomas, Mark~J. Raftery, Cynthia Andrews-Pfannkoch, Matthew Lewis, Jeffrey~M.
  Hoffman, John~A. Gibson, and Ricardo Cavicchioli.
\newblock {Virophage control of antarctic algal host-virus dynamics}.
\newblock {\em {Proceedings of the National Academy of Sciences of the United
  States of America}}, {108}({15}):{6163--6168}, {APR 12} {2011}.

\bibitem{zhou2013}
Jinglie Zhou, Weijia Zhang, Shuling Yan, Jinzhou Xiao, Yuanyuan Zhang, Bailin
  Li, Yingjie Pan, and Yongjie Wang.
\newblock {Diversity of virophages in metagenomic data sets}.
\newblock {\em {Journal of Virology}}, {87}({8}):{4225--4236}, {APR} {2013}.

\end{thebibliography}







\appendix
\section{Appendix}

\subsection{Parameter derivations}
\label{sec:derivation}
\subsubsection{Adsorption rates}
We derive here the rate of viral adsorption.  We assume that absorption and adsorption of the viral particles are diffusion limited and we solve for the rates following \cite{bergpurcell1977}.  By solving an analogous problem of the capacitance of a dielectric sphere coated with conducting disk \textquotedblleft receptor sites" they arrived at the following formulas for a spherical cell's intake of spherical ligands:
\begin{eqnarray*}
J = &2 \pi D a c_{\infty}&
\end{eqnarray*}
where $J$  is the maximum rate of absorption of ligands with diffusion constant $D$ and far-off concentration $c_{\infty}$.  The absorbing spherical cell has diameter $a$.  Our models use the adsorption rate, $\phi$ calculated from $J$:
\begin{eqnarray*}
\phi = \frac{J}{c_{\infty}} =  2 \pi D a,
\end{eqnarray*}
We use the maximum adsorption rate since we sample above and below our reference point.  To estimate $D$ we use the Stokes-Einstein Relation: $D = \frac{k_b T}{3 \pi \eta d}$.  This relation is relevant for spherical particles in low reynolds number fluids which is typical at micro-organismal length scales.  We obtain:

\begin{eqnarray*}
\phi =  \frac{2 a k_b T}{3 \eta d},
\end{eqnarray*}
for a spherical molecule with diameter $d$ in a fluid with viscosity $\eta$ at temperature $T$ where $k_b$ is the Boltzmann constant.  The relevant reference parameters in the units we used for our model are given in Table \ref{table:biophys}.

\begin{table}[h!]
\resizebox{\textwidth}{!}{
\begin{tabular}{ c | l | c | c | c }
\hline
Parameter & Meaning & Value & units & Reference\\
  \hline
  $T$ & Temperature & 293 & K & - \\
  $k_b$ & Boltzmann cons. & $1.0306 \times 10^{-9} $  & $\frac{{cm}^2 kg }{K*{day}^{2}}$ & - \\  
  $\eta$ & Seawater Dynamic Viscosity & $.93312$ & $\frac{kg}{cm*day}$ & \cite{seawater}\\
   $a$ & host diameters & $15*10^{-4}$ (amoeba) $3*10^{-4}$ (cafeteria) & $cm$ & \cite{massana2007} \\
      $d_v$ & virus diameters & $7.5 * 10^{-5}$(mimi) $3 * 10^{-5}$(CroV) & $cm$ & \cite{xiao2005,fischer2010} \\
            $d_p$ & virophage diameters & $7.4* 10^{-6}$ (Sputnik) $6*10^{-6}$ (Mavirus) & $cm$ &  \cite{sun2010,fischer2011}\\
\end{tabular}
}
\caption{Biophysical parameters for determining adsorption/absorption coefficients.}\label{table:biophys}
\end{table}

\subsubsection{Host death rate, d}

We chose our death rate so that the host population will grow to the carrying capacity, $K$, when the viral particles are absent.  The dynamics in this case are:

\begin{eqnarray*}
\dot{H}=  H(b-d(1+\frac{H}{K})),
\end{eqnarray*}
The steady-state host population is:
\begin{eqnarray*}
H^*=  K(\frac{b}{d}-1)
\end{eqnarray*}
Hence, we choose a death rate half the value of the birth rate, $d= \frac{b}{2}$.

\subsubsection{Virophage burst size, $\beta_p$ in PEM}

This value was suggested (by M. Fischer) by counting particle ratios on the electron micrographs in \cite{gaia2013} and \cite{desnues2010}.

\subsection{Reduced model of viral/host abundance}
\label{sec:lumpedmodel}
Here we show that including the virophage in either the IEM or the PEM effectively reduces the burst size of the virus.  For the IEM we define:

\begin{eqnarray*}
H_{\text{total}} &=& H+H_p\\
V_{\text{total}} &=& V\\
\bar{\beta_v} &=& \frac{\beta_v H + \beta_{vp} H_p}{H+H_p}
\end{eqnarray*}
Then we have that
\begin{eqnarray*}
\dot H_{\text{total}} &=& \dot H+\dot H_p \\
&=& H\left[b-d\left(1+\frac{H+H_p}{K}\right)\right] + (1-\rho) b H_p - (\phi_p P +\phi_v V) H + \\
&&H_p\left[\rho b-d\left(1+\frac{H+H_p}{K}\right)\right] +\phi_p H P - \phi_v V H_p \\
  &=& H_{\text{total}} \left[b-d\left(1+\frac{H_{\text{total}}}{K}\right)\right] - \phi_v V_{\text{total}} H_{\text{total}}\\
\dot V_{\text{total}} &=& \dot V = (\beta_v H + \beta_{vp} H_p)\frac{H_{\text{total}}}{H_{\text{total}}} \phi_{v} V-m_v V\\
 &=& \bar{\beta_{v}} \phi_{v} H_{\text{total}} V_{\text{total}} - m_v V_{\text{total}},
\end{eqnarray*}
Thus for the IEM the effective dynamics for the lumped host and viral populations can be thought of as a predator-prey equations with a density dependent viral burst size, $\bar{\beta_v}$.  We assume a negative effect of virophage on viral burst size such that $\beta_{vp}<\beta_v$, which gives
\begin{eqnarray*}
\bar{\beta_v} &=& \frac{\beta_v H + \beta_{vp} H_p}{H+H_p}<\frac{\beta_v H + \beta_{v} H_p}{H+H_p}=\beta_v
\end{eqnarray*}
For the PEM define:

\begin{eqnarray*}
H_{\text{total}} &=& H\\
V_{\text{total}} &=& V+V_p\\
\bar{\beta_v} &=& \frac{\beta_v V + (\beta_{vp}+\beta_i) V_p}{V+V_p}
\end{eqnarray*}
where the infected burst size contribution includes two parameters because viruses are part of the composite relevant to $\beta_i$.  Again, we reduce the dynamics:

\begin{eqnarray*}
\dot H_{\text{total}} &=& \dot H \\
&=& H\left[b-d\left(1+\frac{H}{K}\right)\right] - \phi_v (V+V_p) H \\
&=& H_{\text{total}}\left[b-d\left(1+\frac{H_{\text{total}}}{K}\right)\right] - \phi_v V_{\text{total}} H_{\text{total}}\\
\dot V_{\text{total}} &=& \dot V+\dot V_p \\
 &=& (\beta_v V + \beta_{vp} V_p) \phi_v H - \phi_{vp} V P+m_p V_p - m_v V + \phi_vp V P - (m_p+m_v) V_p + \beta_i \phi_v V_p H\\
&=&(\beta_v V+ (\beta_{vp} + \beta_i) V_p) \frac{V_{\text{total}}}{V_{\text{total}}} \phi_v H - m_v (V+V_p)\\
 &=& \bar{\beta_{v}} \phi_{v} H_{\text{total}} V_{\text{total}} - m_v V_{\text{total}}
\end{eqnarray*}
Thus for both models, the dynamics for the total populations of hosts and viruses reduce to typical predator-prey dynamics with different dynamic burst sizes.  An equilibrium solution to this predator prey dynamical system reveals the virophage effect on host and virus populations in the full form model:

\begin{eqnarray*}
H^*_{\text{total}} &=& \frac{m_v}{\phi_v \bar{\beta^*_v}}\\
V^*_{\text{total}} &=& \frac{1}{\phi_v} \left[b-d\left(1+\frac{m_v}{\phi_v \bar{\beta^*_v}K}\right)\right]
\end{eqnarray*}
where $\bar{\beta^*_v}$ is evaluated with the respective equilibrium populations determined from the full models.  Since both the IEM and PEM were reduced to the same form, this equilibrium condition holds for both models.  The respective boundary equilibrium follows the same form, but with different burst size parameters:

\begin{eqnarray*}
H^*_{b} &=& \frac{m_v}{\phi_v \beta_v}\\
V^*_{b} &=& \frac{1}{\phi_v} \left(b-d\left(1+\frac{m_v}{\phi_v \beta_vK}\right)\right)
\end{eqnarray*}
Since $\beta_{vp} \le \beta_{v}$ and $\beta_{vp}+\beta_i \le \beta_{v}$ , we have $H^*_{\text{total}} \ge H^*_b$ and $V^*_{\text{total}} \le V^*_b$.  This explains why all of our stable coexistence points had higher host densities and lower virus densities when compared to their respective boundary point.

\subsection{Parameter values for figures}
Tables~\ref{table:vfigparams} and~\ref{table:hfigparams} present the parameters used for model simulations in Figure \ref{fig:coexdynamics} and Figure \ref{fig:cycledyn}, corresponding to the PEM and IEM models respectively.

\label{sec:parametervalues}
\begin{table}[h!]
\begin{tabular}{ c | c | c | c | c}
\hline
 Parameter & Fig. \ref{fig:coexdynamics}b & Fig. \ref{fig:coexdynamics}d/\ref{fig:cycledyn}b & Fig. \ref{fig:bistability}a & Fig. \ref{fig:cycledyn}c,d\\   
  \hline   
    $b$ & $1.84$ & $1.53$ & $1.84$ & $1.50$\\
  $d$ & $0.626$ & $0.723$ & $0.626$ & $0.679$\\
    $K$ & $4.32*10^{6}$ & $3.00*10^6$ & $4.32*10^6$ & $2.10*10^6$\\ 
        $\phi_{vp}$ & $1.15*10^{-5}$ & $4.64*10^{-7}$ & $1.15*10^{-5}$ & $1.64*10^{-6}$\\
            $\phi_v$ & $3.79*10^{-6}$ & $3.76*10^{-6}$ & $3.79*10^{-6}$ & $1.98*10^{-5}$\\
$\beta_v$ & $308$ & $134$ & $308$ & $245$\\
    $m_v$ & $0.0269$ & $0.0979$ & $0.0270$ & $0.146$\\ 
      $m_p$ & $0.297$ & $0.0784$ & $0.297$ & $0.0416$\\     
    $\rho_{p}$ & $10.5$ & $2.20$ & $10.5$ & $1.59$\\
    $\rho_{vp}$ & $0.0808$ & $0.588$ & $0.0808$ & $0.386$\\
$\rho_{i}$ & $0.151$ & $0.0778$ & $0.151$ & $0.443$\\  
\end{tabular}
\caption{PEM figure parameter sets shown to 3 significant figures.}
\label{table:vfigparams}
\begin{tabular}{ c | c | c}
\hline
 Parameter & Fig. \ref{fig:coexdynamics}a & Fig. \ref{fig:coexdynamics}c/\ref{fig:cycledyn}a\\
  \hline
    $b$ & $1.99$ & $1.15$ \\
  $d$ & $0.862$ & $0.913$ \\ 
  $K$ & $4.61*10^{6}$ & $7.69*10^6$ \\
      $\phi_p$ & $5.51*10^{-6}$ & $1.53*10^{-6}$ \\
          $\phi_v$ & $1.77*10^{-6}$ & $8.81*10^{-6}$\\
   $\beta_v$ & $157$ & $162$\\    
    $m_v$ & $0.0646$ & $0.567$\\ 
      $m_p$ & $0.274$ & $0.145$\\ 
    $\rho_p$ & $13.8$ & $1.12$\\ 
    $\rho_{vp}$ & $0.343$ & $0.348$\\
    $\rho$ & $0.392$ & $.860$ \\ 
\end{tabular}
\caption{IEM figure parameter sets shown to 3 significant figures.}
\label{table:hfigparams}
\end{table}

\subsection{Phase lag of viruses and virophage differentiate the two models during cyclical dynamics}
\label{sec:cycles}

\begin{figure}[h!]
  (a)
    \includegraphics[width=0.45\textwidth]{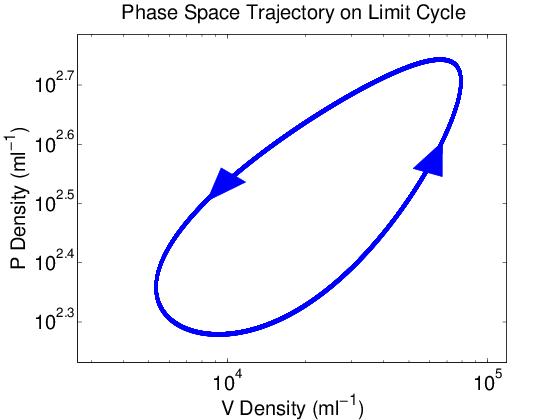}
      (b)
    \includegraphics[width=0.45\textwidth]{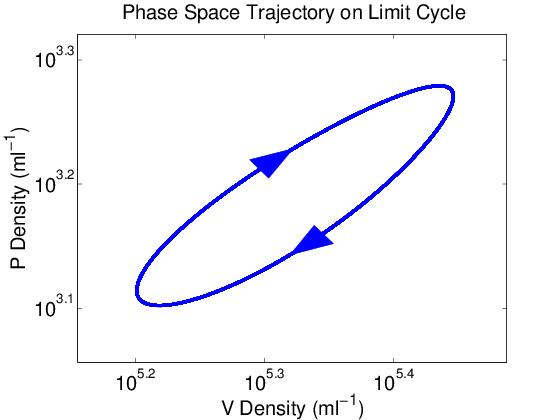}\\
      (c)
    \includegraphics[width=0.45\textwidth]{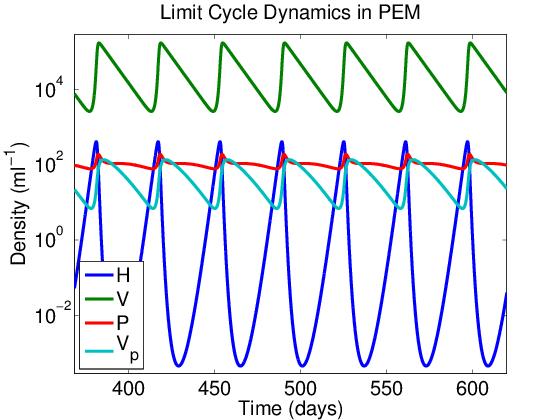}
      (d)
    \includegraphics[width=0.45\textwidth]{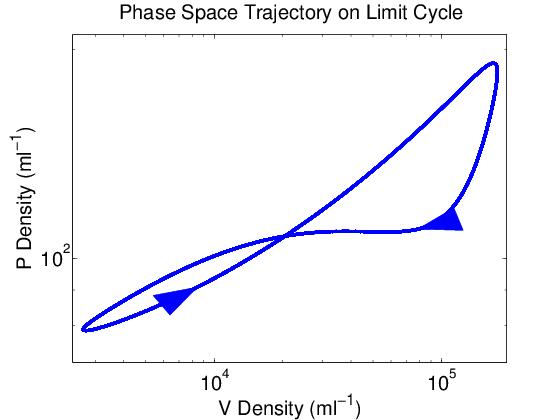}
      \caption{Phase space representation of presented cyclical dynamics in Figures \ref{fig:coexdynamics}b,d for (a) IEM and (b) PEM. (c) Cyclical coexistence dynamics in PEM and respective (d) phase space representation.  The parameter values are shown in Table \ref{table:vfigparams}. \label{fig:cycledyn}}
\end{figure}

The order of virus and virophage peaks during cyclical dynamics may be a means for distinguishing the modes of coinfection from population level data.  Specifically, the direction of the cycles in the V-P phase subspace have been found to be opposite in their orientation.  We use the cyclical dynamics from Figures \ref{fig:coexdynamics}b,d as an example.  The corresponding phase space dynamics for the IEM projected to the V-P subspace are shown in figure \ref{fig:cycledyn}a.  The counterclockwise movement of this phase trajectory corresponds to the virophage population peak lagging the virus population peak.  The phase space dynamics for the PEM projected onto the V-P subspace are shown in figure \ref{fig:cycledyn}b.  The counterclockwise movement of this phase trajectory corresponds to the virus population peak lagging the virophage population peak.  The same analysis was performed on a number of other points with similar results, however due to small number of examples explored, this preliminary result warrants further analysis.  For example, we identified a cycle in the PEM with different dynamics when projected onto the V-P subspace as shown in Figures \ref{fig:cycledyn}c,d.

\end{document}